\documentclass[article]{elsart}
\usepackage{epsfig}
\usepackage{graphicx}
\usepackage{fancybox}

\newcommand{\ds}{{\rm d}}

\newcommand{\ba}{\begin{eqnarray}}
\newcommand{\ea}{\end{eqnarray}}
\newcommand{\be}{\begin{equation}}
\newcommand{\ee}{\end{equation}}

\begin{document}

\begin{frontmatter}
\title{Pattern forming pulled fronts:\\ bounds and universal convergence}
\author{Ute Ebert$^{1,2}$, Willem Spruijt$^3$ and Wim van Saarloos$^3$} 
\address{$^1$ CWI, Postbus 94079, 1090 GB Amsterdam, The Netherlands,}
\address{$^2$ Department of  Physics, TU Eindhoven, Postbus 513, 
  5600 MB Eindhoven, \\The Netherlands,}
\address{$^3$ Instituut--Lorentz, Leiden University, Postbus 9506,
  2300 RA Leiden, \\The Netherlands}
\maketitle
\begin{abstract}
We analyze the dynamics of pattern forming
fronts which propagate into an unstable 
state, and whose dynamics is of the pulled type, so that their
asymptotic speed is equal to the linear spreading speed $v^*$. 
We discuss a method that
allows to derive bounds on the front velocity, and which hence can be
used to prove for,  among others,   the Swift-Hohenberg equation, the Extended
Fisher-Kolmogorov equation and the cubic Complex Ginzburg-Landau equation,
that the dynamically relevant fronts are of the pulled type.
In addition, we generalize the derivation of the universal 
power law convergence of
the dynamics of uniformly translating
pulled fronts to both coherent and incoherent
pattern forming fronts. The analysis is based on a matching analysis 
of the dynamics in
the leading edge of the front, to  the behavior imposed by the
nonlinear region behind it. Numerical simulations of fronts 
in the Swift-Hohenberg
equation are in full accord with our analytical predictions. 

\end{abstract}
\end{frontmatter}

 
\section{Introduction}

In the last few years, it has become clear that when considering
a problem of a front  which propagates into an unstable state, 
it is crucial to
distinguish two different classes, according to whether their
asymptotic speed is equal to or larger than the linear spreading speed
$v^*$. The linear spreading speed is  a simple concept that dates back to
developments in plasma physics and fluid dynamics that took place
almost half a century ago \cite{briggs,bers,ll,huerre}. It is the
asymptotic speed with which an initially localized perturbation
 about the unstable state
spreads into this unstable state according to the {\em linear}
dynamics, the dynamics obtained by linearizing the dynamical equations
about the unstable state. For any deterministic dynamical equation
this linear spreading speed $v^*$ can be determined explicitly from a
long-time asymptotic saddle-point type analysis of the Green's function 
of the relevant
dynamical equation.  In practice, therefore, $v^*$ is given explicitly
by the dispersion relation of Fourier modes obeying the linearized
dynamical equation \cite{briggs,bers,ll,huerre,evs1,evs2,wimreview}.

Given the existence of a finite linear spreading speed $v^*$ for a
given problem, only  two different types of asymptotic
front solutions can emerge starting from ``steep'' or ``sufficiently
localized'' initial conditions: either the asymptotic velocity of the
nonlinear front is {\em equal to} $v^*$ or it is {\em larger than}
$v^*$. In the first case we speak of ``pulled fronts'',  as such fronts
are essentially being pulled along by the growth and spreading of the
linear dynamcs  in the leading edge where the linearized dynamical
equations can be used. In the second case of fronts whose
asymptotic speed is larger than $v^*$, we speak of pushed fronts
\cite{evs1,evs2,wimreview,stokes,paquette}.  Because the essential dynamics of pulled
fronts is actually taking place in the region {\em ahead of } the
nonlinear front region, their properties are very different from
pushed fronts or other fronts, domain walls or kink solutions whose
properties are determined by a nonlinear eigenvalue problem: the
 singular perturbation theory which is normally used to map weakly
curved fronts onto a moving boundary problem, breaks down for pulled
fronts \cite{evs3}, and their velocity and shape converge with universal 
power laws
to their asymptotic value and shape. For nonlinear diffusion equations of the
type studied by Fisher \cite{fisher} and Kolmogorov {\em et al.}
\cite{kpp}  the first term expressing this power law
convergence was already derived in 1983 by  Bramson
\cite{bramson}, but we have recently found that this slow power law
convergence can be summarized in one single exact equation that governs any
pulled front which converges to a uniformly translating solution
\cite{evs1,evs2,evsp}.  For a review of many of these results, 
see \cite{wimreview}.

\begin{figure}[t]
\begin{center}
\hspace*{-4mm}
\epsfig{figure=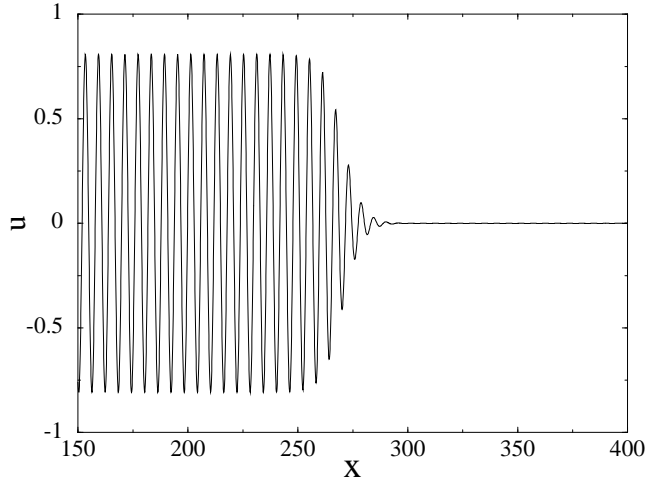,angle=-90,width=0.6\linewidth}
\end{center}
\caption[]{Snapshot of a front in the Swift-Hohenberg equation 
(\ref{swifthohenberg}) for
$\varepsilon=0.5$. The front propagates to the right into the region where
$u$ is in the unstable state $u=0$.   }\label{figprofile}
\end{figure}

As it turns out, the matching analysis on which the derivation of the
power law convergence  is based (see also \cite{bd}), requires only minimal input on the
form of the nonlinear uniformly translating pulled front solution to
which the front solution converges --- the explicit expressions for
the velocity convergence are all obtained from a proper Ansatz for the
asymptotic expansion of the front solutions in the leading edge, the
region where the dynamical equation can be linearized. It is the
purpose of this paper to show that this part of the analysis can be
easily generalized to dynamical equations whose dynamics is pattern
forming, i.e., whose asymptotic front solutions are {\em not} 
uniformly translating. 
An example of such a front in the Swift-Hohenberg equation is shown
in Fig.~\ref{figprofile}. This conclusion was already announced without
derivation in  \cite{storm}.  In fact, the asymptotic relaxation
formula which we derive here also applies to incoherent pattern
forming fronts --- the reason is that the linear spreading dynamics is
always coherent, irrespective of whether the dynamics in the nonlinear
region behind the front is coherent or incoherent
\cite{wimreview,storm}. While the asymptotic expansion in the leading
edge which we will discuss here thus pertains to both types of fronts,
we shall focus our discussion of the application of the formula on coherent
pattern forming fronts.  

One of the simplest examples of a dynamical
equation whose pattern forming fronts are coherent is the
Swift-Hohenberg equation, and we will therefore use this equation to
illustrate and test our analytical results. In fact, the
Swift-Hohenberg equation has often played a role in studies of front
propagation \cite{dee,vs2,collet1,collet2,eckmann,collet3} --- it is
essentially the only equation with pattern 
forming fronts for which a number of exact results (including the convergence
to a pulled front solution)  are known
\cite{collet1,collet2,eckmann,collet3}. 

Because there are so few
rigorous results for pattern forming fronts in general, we will, before turning
to the analysis of the front convergence, discuss a method which
allows us to derive a bound on the velocity for pattern forming fonts, like the
Swift-Hohenberg equation, the Extended Fisher-Kolmogorov equation, or the
cubic Complex Ginzburg-Landau equation.
Although our  argument is in essence a simplified version of
the line of analysis Collet
and Eckmann \cite{collet3} use to prove that fronts in the Swift-Hohenberg
equation are pulled, we do want to show the reader how in just a few lines one
can prove that fronts in pattern forming equations are pulled: we think that the method holds
the promise for many new rigorous results on front propagation.

The layout of this paper is as follows. In the next section we first
discuss our method to derive a bound on the front velocity. Then, 
in section III we
perform the asymptotic expansion of the dynamics of the leading edge
of a pattern forming front,
which gives the expressions for the convergence of the front velocity
and shape to their asymptotic behavior. In section IV we illustrate
these results  with numerical solutions of   the Swift-Hohenberg
equation, and we close the paper with a brief summary.

\section{The linear spreading velocity as a rigorous upper bound}

\subsection{The linear spreading velocity}

We consider a generic dynamical equation
for some generic dynamical variable $\phi$, whose stationary state $\phi=0$ is
linearly unstable, and whose dispersion relation is given by $\omega(k)$. 
This means that a Fourier perturbation $e^{ikx}$ of the unstable state 
evolves under the linear
dynamics as $e^{-i\omega(k)t+ikx}$. Associated with the linear
dynamical problem is a {\em linear spreading velocity} $v^*$, the
velocity with which an initially localized perturbation spreads
asymptotically into the unstable state 
according to the linearized dynamics. The asymptotic spreading is
simply determined by a long-time saddle point analysis of the Green's 
function of the linear equation. From this analysis, one finds $v^*$
explicitly in terms of $\omega(k)$ as 
\cite{briggs,bers,ll,huerre,evs1,evs2,wimreview}
\begin{equation} 
\left. \frac{\ds \omega(k)}{\ds  k}\right|_{k^*} = 
\frac{{\rm Im}~\omega(k^*) }{\lambda^*} ,
\hspace*{0.9cm}
 v^* =   \frac{{\rm Im}~\omega(k^* )}{\lambda^*} , 
\hspace*{0.9cm} k^*\equiv q^*+i\lambda^* .\label{saddlepoint}
\end{equation}
The first equation determines the saddle point value $k^*$ in the
complex plane, and the second one then gives the linear spreading velocity
$v^*$. The third equation fixes our notation for the splitting of $k^*$ 
into real part $q^*$ and imaginary part $\lambda^*$ for the remainder 
of the paper. The complex parameter  
\begin{equation}
{\mathcal D} \equiv \left. \frac{i}{2} \frac {\ds ^2\omega(k)}
{\ds k^2}\right|_{k^*} \label{Ddefinition}
\end{equation}
plays the role of a complex diffusion coefficient\footnote{${\mathcal D}$
is the complex generalization of the diffusion constant $D$ as 
in \cite{wimreview} and should not be confused
with the operator defined in Eq.~(5.27) of \cite{evs2}.}.
If there are several saddle points, the one with the largest $v^*$ 
is the relevant one \cite{evs2}. Since the growth rate 
Im~$(-i\omega(k)+iv^*k)$ in the comoving frame is maximal at
$k^*$ for a relevant saddle point, the sign of Re~${\mathcal D}$
is fixed: 
\be
{\rm Re}~{\mathcal D}>0.\label{Deq}
\ee
Our analysis applies to sufficiently steep initial conditions \cite{evs2} 
\be
\label{steep}
\lim_{x\to\infty}\phi(x,0)\;e^{\lambda x}=0\qquad\mbox{for some }
\lambda>\lambda^*;
\ee
initial conditions with bounded support fall into this class.
An important result 
is that in a frame $\xi=x-v^*t$ moving with velocity $v^*$ to the right, 
the asymptotic evolution of the field under the linear equation is given by
\be
\label{phibasic}
\phi (x,t) \sim
e^{\textstyle-\lambda^*\xi+iq^*\xi-i\Omega(k^*)t}~~
\frac{e^{\textstyle -\xi^2/4{\mathcal D}t}}{\sqrt{4\pi {\mathcal D} t}},
\ee
where
\be \Omega(k)\equiv\omega(k)-v^*k,
\ee
and where the co-moving coordinate
\be  \xi=x-v^*t 
\ee
is held  fixed while $t\to \infty$.
This follows from the saddle point analysis of the Green's function 
in the limit of large $t$, cf.~sections 5.3 and 5.5.1 in \cite{evs2}.
The saddle point equations (\ref{saddlepoint}) can be expressed in terms of 
$\Omega(k)$ as $\ds_k\Omega|_{k^*}=0$ and Im~$\Omega(k^*)=0$.
For the remaining real part of $\Omega(k^*)$, we use the notation 
\be
\Omega^*=\Omega(k^*)=\mbox{Re}\;\Omega(k^*).
\ee

Eq.\ (\ref{phibasic}) illustrates that an initially sufficiently localized 
linear perturbation reaches
the velocity $v^*$ and the spatial decay rate $\lambda^*$ for $t\to\infty$
under the dynamics of the linearized equation.

\subsection{Upper bounds on the velocity: proof of pulling} 

When a front evolves 
under the full nonlinear equation into an unstable state, 
its asymptotic speed can never be smaller than the linear spreading 
velocity $v^*$.
If the asymptotic speed equals $v^*$, the front is called pulled
\cite{evs1,evs2,wimreview,stokes}, otherwise it is called pushed. As
a rule of thumb, dynamical equations whose nonlinear terms are all
suppressing the growth lead to pulled fronts, but there is at present
no general theory that allows one to predict when fronts are pulled
and when they are pushed.

In the present section, a simple proof is given that fronts in some
pattern forming equations are pulled.

\subsubsection{A real field $\phi$ with nonlinearity 
${\mathcal N}(\phi)\;\phi$}

Consider first an equation of motion for a real field $\phi$ 
\ba
\label{con1}
\partial_t\phi=\sum_{n=0}^Na_n \partial_x^n\phi - {\mathcal N}(\phi)\;\phi
,~~~~~{\mathcal N}(0)=0,
\ea
with explicit linear terms and a nonlinearity\footnote{Note that 
in \cite{evs2} the complete nonlinear expression 
${\mathcal N}(\phi)\;\phi$ was denoted as $N(\phi)$, but the present notation
turns out to be more convenient for the generalizations.} ${\mathcal N}(\phi)\;\phi$.
Examples of such equations are the nonlinear diffusion equation $\partial_t u= \partial^2_xu +f(u)$,
the Swift-Hohenberg-equation
\begin{equation}
\partial_t u  = \varepsilon u  - (\partial^2_x+1)^2 u  -u^3 =
(\varepsilon-1) u  - 2 \partial^2_x u  -\partial^4_x u - u^3,
\label{swifthohenberg} ~~ (\varepsilon >0),
\end{equation}
or the Extended Fisher-Kolmogorov (EFK) equation \cite{wimreview,deevs,bert}
\begin{equation}
\partial_t u  =     \partial^2_x u  -\gamma \partial^4_x u  + u- u^3.
\label{efk} 
\end{equation}
The linear operator determines the dispersion relation $\omega(k)$
and the parameters $v^*$, $\Omega^*$, $q^*$, $\lambda^*$ and ${\mathcal D}$ 
as discussed above.

The relevant dynamics of a pulled front that leaves
a homogeneous state behind ($\Omega^*=0=q^*$), was identified in \cite{evs2}
by the leading edge transformation $\phi(x,t)=e^{-\lambda^*\xi}\psi(\xi,t)$,
$\xi=x-v^*t$. For pattern forming fronts with
$\Omega^*\ne0\ne q^*$, different generalizations of this transformation
are possible. While in the next section dealing with the asymptotic dynamics,
the complete complex phase $e^{-\lambda^*\xi+iq^*\xi-i\Omega^*t}$
will be factored out of $\phi$, for deriving bounds, 
it will be more convenient here to factor out the envelope 
$e^{-\lambda^*\xi}$.
In a frame moving with velocity $v^*$, the field $\hat\psi(\xi,t)$
is then defined through
\ba
\label{genPhi}
\phi(x,t)=e^{-\lambda^*\xi}\hat\psi(\xi,t),~~~~~~\xi=x-v^*t.
\ea
The effect of the transformation is demonstrated by comparing 
Fig.~\ref{figprofile} with Fig.~\ref{fig1} below
which show the original dynamical field $u$ 
of the Swift-Hohenberg equation and the associated field $\hat\psi$. 
The field $\hat\psi$ in Fig.~\ref{fig1}
magnifies the relevant dynamics in the leading edge which we
will analyze in section 3, while this dynamics
is hidden in Fig.~\ref{figprofile}.

With this transformation, the equation of motion for $\hat\psi$ becomes
\ba
\label{con2}
\partial_t\hat\psi
-v^*\left(\partial_{\xi}-\lambda^*\right)\hat\psi
=\sum_{n=0}^N
a_n\left(\partial_{\xi}-\lambda^*\right)^n\hat\psi 
- {\mathcal N}\left(\hat\psi\;e^{-\lambda^*\xi}\right)\;\hat\psi.
\ea
With the two auxiliary functions of the Fourier variable $k$
\ba
\label{con2a}
\sigma(k)&=&\sum_{n=0}^N a_n (ik-\lambda^*)^n+v^*(ik-\lambda^*)
=-i\omega(k+i\lambda^*)+iv^*(k+i\lambda^*),\nonumber\\
\bar\psi(k,t)&=&\int_{-\infty}^\infty {\rm d}\xi
\;\hat\psi(\xi,t)\;e^{-ik\xi},
\ea
the linear operators in Eq.\ (\ref{con2}) 
can be written in a more compact form
\ba
\label{con3}
\partial_t\hat\psi(\xi,t)=\int_{-\infty}^\infty
\frac{ {\rm d} k}{2\pi} \;e^{ik\xi}\;\sigma(k)\;\bar\psi(k,t) 
- {\mathcal N}\left(\hat\psi\;e^{-\lambda^*\xi}\right)\;\hat\psi.
\ea
Now multiply the equation with $\hat\psi(\xi,t)$ and integrate
over space. Using the identity
\ba
\int {\rm d} k\;\sigma(k)\;\bar\psi(k,t)\;\bar\psi(-k,t)
= \int {\rm d} k\;{\rm Re}~\sigma(k)\;\left|\bar\psi(k,t)\right|^2,
\ea
the final result is
\ba
\label{con4}
\frac{\partial}{\partial t} \int {\rm d}\xi\;\frac{\hat\psi^2(\xi,t)}{2}&=&
 \int \frac{ {\rm d} k}{2\pi} \;{\rm Re}~\sigma(k)\;\left|\bar\psi(k,t)\right|^2
\nonumber\\ 
&&
-\int {\rm d}\xi\; {\mathcal N}\left(\hat\psi(\xi,t)\;e^{-\lambda^*\xi}\right)
\;\hat\psi^2(\xi,t).
\ea
If $\phi$ initially is sufficiently steep (\ref{steep}) for $x\to\infty$,
and if $|\phi|$ stays bounded behind the front at $x\to-\infty$, 
then the integrals exist initially.
If furthermore the right hand side of (\ref{con4}) can be shown
to be negative and of order $\int d\xi \;\hat\psi^2$, 
then $\int {\rm d}\xi\;\hat\psi^2(\xi,t)\downarrow 0$ for growing $t$.
This means that in a frame moving with velocity $v^*$, $\hat\psi^2$
vanishes; and this implies that the front cannot move faster
than $v^*$ for $t\to\infty$.

For the r.h.s.\ of (\ref{con4}) to be negative, we need both 
integrals to be negative. Since Re~$\sigma(k)=
\mbox{Im}~\omega(k+i\lambda^*)-v^*\lambda^*$,
the saddle point construction entails that
Re~$\sigma(q^*)=0$, $\partial_k\sigma|_{q^*}=0$ and
$\partial_k^2\sigma|_{q^*}=-2{\mathcal D}$ with Re~${\mathcal D}>0$.
Therefore
\be
{\rm Re}~\sigma(k)\le0~~~\mbox{for all real }k.  \label{sigma}
\ee
If there are several saddle points, this condition holds for the one 
corresponding to the largest spreading speed $v^*$ \cite{evs2}. The present
formulation in terms of $\sigma(k)$  yields another route 
to  this conclusion.

The sign of the integral over the nonlinearity is
fixed if the sign of ${\mathcal N}$ is fixed. Therefore 
a sufficient criterion for the front to be pulled is
\be
{\mathcal N}(\phi)\ge 0~~~\mbox{for all relevant }\phi.
\ee
In a pattern forming front, the sign of $\phi$ can change.
This increases the relevant values of $\phi$ and therefore decreases
the admissible functions ${\mathcal N}$. E.g., for 
${\mathcal N}(\phi)=\phi^r$, a monotonic front with non-negative $\phi$
will be certainly pulled for all $r>0$, while for a pattern forming
front, $r$ needs to be an even integer.
Both in the Swift-Hohenberg and EFK equation, ${\mathcal N}$ is quadratic in
the dynamical variable, hence the above argument immediately shows that sufficiently
steep initial conditions lead to pulled fronts in these equations. With a few slight
modifications, the analysis can also be extended to the difference equation
$dC_i/dt= C_i- C^2_{i-1}$, for which fronts were empirically found to be pulled \cite{evsp,ramses}.

\subsubsection{A complex field $A$: the complex Ginzburg-Landau-equation}
It was already remarked by Collet and Eckmann in a footnote in
\cite{collet3} that the above line of analysis can be 
extended to the case of the cubic Complex Ginzburg Landau equation. 
We present the argument here in our language, and then generalize it 
to an even more general class of
equations in the next subsection.

We analyze the complex Ginzburg-Landau-equation for complex field
$A(x,t)$
\be
\label{CGL}
\partial_tA=\epsilon A+(1+c_1)\partial_x^2 A-(1-ic_3)|A|^2A
,~~~~~\mbox{with }\epsilon,c_1,c_3 \mbox{ real},
\ee
or more generally an equation of the form
\be
\label{genA}
\partial_tA=\sum_{n=0}^N a_n \partial_x^n A - {\mathcal N}(A)\;A
,~~~~~\mbox{with }A(x,t),~a_n \mbox{ complex}.
\ee
The saddle point parameters $\lambda^*$, $q^*$, $v^*$, $\Omega^*$
and ${\mathcal D}$ are again used for the transformation
\be 
\hat\psi(\xi,t)=e^{-\lambda^*\xi}\;A(x,t),~~~~~
\mbox{where }\xi=x-v^*t.
\ee
The calculation now follows essentially the lines of the previous 
calculation --- except that one has to take into account that 
the field $\hat\psi$ and the coefficients are now complex. 
Therefore the equations of motion for $A^*$ or $\hat\psi^*$
have to be considered, too. They are, of course, derived
by simply taking the complex conjugate of the equations 
for $A$ and $\hat\psi$. One then easily derives an equation
for $\hat\psi^*\partial_t\hat\psi++\hat\psi\partial_t\hat\psi^*=
\partial_t|\hat\psi|^2$ that after spatial integration
and a few steps of calculation can be reduced to
\be
\frac{\partial}{\partial t}\int {\rm d} \xi\;\frac{|\hat\psi(\xi,t)|^2}{2}
=\int \frac{ {\rm d}k}{2\pi}\;\mbox{Re}~\sigma(k)\;|\bar\psi(k,t)|^2
-\int d\xi\;\mbox{Re}~{\mathcal N}(A)\;|\hat\psi|^2.
\ee
Here $\bar\psi(k,t)$ and $\sigma(k)$ are defined precisely 
as in (\ref{con2a}).

This means that the complex equation has been reduced to expressions
that contain absolute values and real parts only. Therefore the 
conclusion from the previous subsection is easily extended:
an equation of form (\ref{CGL}) or (\ref{genA}) creates pulled fronts if
\be
\mbox{Re}~{\mathcal N}(A)\ge0~~~\mbox{for all relevant }A.
\ee

This is a nontrivial result, since in contrast to the
real equation (\ref{genPhi}), the complex equation does not have
an energy minimizing structure;  still the bound 
can be derived in the same way as before. Specialized to the cubic
Complex Ginzburg-Landau equation, the above analysis simply proves that fronts in this
equation are pulled, a fact known already empirically since over 20 years \cite{wimreview,nb}.

\subsubsection{Generalization of admissible linearities and nonlinearities}

In the last step, the admissible linear and nonlinear operators 
are reconsidered and generalized. For complex functions $A$,
the general form is
\be
\label{eq22}
{\mathcal L} A + {\mathcal N}(A,\partial_xA,\partial_x^2A,
\ldots,\partial_x^mA)\;A=0,
\ee
where ${\mathcal N}$ again can be complex.
${\mathcal L}$ is an arbitrary complex linear operator that can take
the differential form above, but also a difference or integral
or mixed form as discussed in Section V of \cite{evs2}. It determines 
the saddle point parameters $v^*$, $\lambda^*$, $q^*$ and ${\mathcal D}$.
Independent of the original functional form of the linear operator,
the expansion about the (large-$t$, large-$x$)-saddle point will
lead to the differential form
\be
\label{eq23}
\tau_0\partial_t\hat\psi = \ldots - {\mathcal N}(A,\partial_xA,\partial_x^2A,
\ldots,\partial_x^mA)\;\hat\psi.
\ee
The analysis now proceeds as before with the final result
\be
\frac{\partial}{\partial t}\int {\rm d} \xi\;
\frac{|\hat\psi(\xi,t)|^2}{2} 
=\ldots
-\int {\rm d} \xi\;\mbox{Re}~\frac{{\mathcal N}(A,\partial_xA,\ldots,\partial_x^mA)}{\tau_0}
\;|\hat\psi|^2.
\ee
A sufficient criterion for the front to be pulled is 
\be
\label{genN}
\mbox{Re}~\frac{{\mathcal N}(A,\partial_xA,\ldots,\partial_x^mA)}
{\tau_0}\ge 0~~~\mbox{for all relevant }A.
\ee

In essence, the method discussed here confirms mathematically what one
would expect intuitively for equations where only the linear terms lead to growth away
from the unstable state $\phi=0$, while all the nonlinear terms are clearly stabilizing. In
such cases, fronts are shown to be of the pulled type. There are several cases where fronts
are empirically known to be pulled, but  where the method in its present formulation fails. E.g., 
while adding a 
nonlinearity like 
$-(\partial_x u)^2u $ to the Swift-Hohenberg equation (\ref{swifthohenberg}) or EFK
equation (\ref{efk}) leaves the fronts in these equations of the pulled type,
since ${\mathcal N}=(\partial_x u)^2\ge0$,
the nonlinearity of the Kuramoto-Sivashinsky-equation 
$\partial_t u=-\partial_x^2 u-\partial_x^4u +(\partial_x u)u $ 
does not fall into the class (\ref{genN}). In fact, extending the method to the Kuramoto-Sivashinsky equation must clearly be quite a challenge, since adding a linear term $c \partial^3_x u$ 
gives a transition to pushed fronts for $c\approx 0.15$ \cite{wimreview}.
An  easier challenge to start with appears to be the
 the Cahn-Hilliard-equation
$\partial_t u=-\partial_x^2\left(\partial_x^2 u +u -u ^3\right)$. 
Again, in its present form
our method does not apply straightforwardly to the Cahn-Hilliard equation.
Nevertheless, for a front penetrating the state $u =0$ 
under the Cahn-Hilliard dynamics, 
we derive after a few partial integrations that
\be
\partial_t\int {\rm d} \xi\;\hat\psi^2=\ldots-3 \int {\rm d} \xi\;\hat\psi^2
\left(\left(\partial_x u \right)^2-\left(\lambda^*u \right)^2\right).
\ee
It is very likely that the sign of this integral over the
nonlinearity is negative, since $(\partial_x u )/ u $ is the local slope
of the full oscillating front, while $(\lambda^*u )/u $ is the slope
of only the envelope in the leading edge. However, we have not yet been able
to prove this.

In summary, we have derived sufficient criteria for a large class
of equations to form pulled fronts, i.e., fronts that propagate with
the linear spreading speed $v^*$. We now proceed to determining their 
actual rate of convergence to the asymptotic behavior.

\section{Power law convergence to the asymptotic speed and shape of a
pulled front}

In \cite{evs2} we have analyzed pulled fronts that for long times
approach uniformly translating fronts, and we have derived their
rate of convergence to the asymptotic velocity and front profile.
We will now extend this analysis to pattern forming fronts.

Our analysis in \cite{evs2} was based on a complete matching 
of the transient dynamics in the leading edge (where the nonlinearities 
in the dynamical equation can be neglected) to the behavior in 
the nonlinear front region itself.  This detailed analysis explicitly 
demonstrates that the matching procedure can be carried out order by order. 
It is remarkable and in line with the picture that has emerged 
for the pulled front mechanism, that the
coefficients in the asymptotic expressions are actually
obtained from the asymptotic analysis in the leading edge only;
more precisely they are given by the saddle point parameters 
(\ref{saddlepoint}), (\ref{Ddefinition}) of the linearized equation. This
is because  for the analysis in the leading edge only input on the dominant 
analytic behavior of the asymptotic front profile is needed\footnote{
  In the language of a matching analysis, the outer (leading edge)
  expansion of the inner (nonlinear front) solution is expressed by
  the condition (\ref{matchingcond}) below.}. 
For brevity, we will therefore present
here only the generalization of the asymptotic expansion in the
leading edge, following the lines of our earlier paper. 

\subsection{The dynamical equation for the leading edge variable
$\psi$ in the frame $\xi_X$}

The first ingredient of the asymptotic analysis for the front 
convergence is to note that in the leading edge, the saddle point 
analysis from Section 2.1 implies that the field $\psi(\xi,t)$
defined through
\begin{equation}
\phi(x,t)  = e^{-\lambda^*\xi}\;e^{iq^*\xi -i\Omega^*t} \;\psi(\xi,t)
,~~~~\xi=x-v^*t .
\label{leadingedgetrafo}
\end{equation}
becomes a function which varies slowly in space and time for large $x$ 
and $t$, and this slow dynamics is governed by a generalized diffusion 
equation of the form
\begin{equation}
\frac{\partial \psi}{\partial t} =  {\mathcal D} \frac{\partial^2
\psi}{\partial \xi^2}  + {\sf {\mathcal D}_3} \frac{\partial^3\psi}{\partial
\xi^3} + w \frac{\partial^2 \psi}{\partial t \partial \xi} + \tau_2
\frac{\partial^2 \psi}{\partial t^2} + \cdots -{\mathcal N}(\phi,\ldots)\;\psi.
\label{leadingedge2}
\end{equation}
In the function $\psi$, the full complex prefactor is factorized
out of $\phi$, in contrast to the partial factorization in Eq.\ (\ref{con2}). 
The parameter ${\mathcal D}$ is the generalized diffusion coefficient defined 
already in  Eq.~(\ref{Ddefinition}) above. Likewise the other
expansion coefficients ${\mathcal D}_3, w, \tau_2$ et cetera
can all be expressed in terms of the expansion of the dispersion
relation near the saddle point --- see Eq.~(5.64) of 
\cite{evs2}. E.g., we simply have 
 ${\mathcal D}_3= (1/3!)   {\rm d}^3\omega/{\rm d} k^3|_{k^*}$. 
Note that we call Eq.~(\ref{leadingedge2}) a generalized
diffusion equation since the dominant terms for large $\xi$ and $t$
are in fact diffusive and can generate the Gaussian from Eq.~(\ref{phibasic}).

For  equations which lead to uniformly translating fronts, $q^*=0$ 
and ${\mathcal D}$ is real, but for pattern forming fronts
${\mathcal D}$ is generally complex and   $ q^*\neq 0$. 

As discussed in \cite{evs1,evs2,wimreview}, if we follow a level line where 
$|\phi|$ is constant, the $1/\sqrt{t}$ term in (\ref{phibasic})   
implies an {\em unbounded} logarithmic shift in the position of the
level 
line, and hence of the transient fronts in the nonlinear equation. 
 The crux of the convergence analysis is therefore to introduce a
 collective coordinate $X(t)$ for the front position,    
\begin{equation}
\dot{X}(t)= \frac{c_1}{t} + \frac{c_{3/2}}{t^{3/2} } + \frac{c_2}{t^2} +
\cdots  ~~~\Longleftrightarrow ~~X(t)=c_1 \ln t - \frac{2
  c_{3/2}}{t^{1/2}} +\cdots, \label{Xeq}
\end{equation}
and to perform an expansion in the {\em logarithmically shifted frame}
\begin{equation}
\xi_X= \xi-X(t) = x-v^*t -X(t).  
\end{equation}

For pattern forming fronts, we likewise introduce a global  
time-dependent phase $\Gamma(t)$,
\begin{equation}
\dot{\Gamma}(t)= \frac{d_1}{t} + \frac{d_{3/2}}{t^{3/2} } + \frac{d_2}{t^2} +
\cdots  ~~~\Longleftrightarrow ~~\Gamma(t)=d_1 \ln t - \frac{2
  d_{3/2}}{t^{1/2}} +\cdots,
\label{Gammaeq}
\end{equation}
and we define the field $\psi_X$ 
in the shifted frame $\xi_X$ and with a global   slow phase factor 
$\Gamma$ by writing $\phi$ as 
\begin{equation}
\phi(x,t)  =  e^{-\lambda^*\xi_x}e^{iq^*\xi_X -i(\Omega^*t+\Gamma(t))}
\;\psi_X(\xi_X,t).\label{leadingedgetrafo2}
\end{equation}
Comparison of (\ref{leadingedgetrafo}) and (\ref{leadingedgetrafo2}) 
shows that 
\begin{equation}
\psi(\xi,t)= e^{\lambda^*X(t)-iq^*X(t)- i\Gamma(t)} \;\psi_X(\xi_X,t).
\end{equation}
With this transformation, we obtain from (\ref{leadingedge2})  the
relevant dynamical
equation\footnote{The term proportional to $w$ is not present for
equations like the Swift-Hohenberg 
equation or for the Complex Ginzburg Landau equation, but can be
present in more general cases. 
As was already found for uniformly translating fronts \cite{evs2}, 
this term does not affect the relevant 
terms for the power law relaxation.}  for $\psi_X(\xi_X,t)$  
\begin{eqnarray}\nonumber 
\frac{\partial \psi_X}{\partial t} &- & 
\dot{X}(t)\left(ik^*+ \frac{\partial}{\partial \xi_X} \right) \psi_X 
-  i\dot{\Gamma} 
(t) \psi_X =  {\mathcal D} \frac{\partial^2
\psi_X}{\partial \xi_X^2}  + {\sf {\mathcal D}_3} \frac{\partial^3\psi_X}{\partial
\xi_X^3}+\ldots \\  & & \hspace*{0.45cm} + w \left[ \frac{\partial}{\partial t
} - \dot{X}(t) 
\left(ik^*+ \frac{\partial}{\partial \xi_X} \right) - i\dot{\Gamma}(t)
\right] \frac{\partial \psi_X}{\partial \xi_X} + \cdots 
-{\mathcal N}\;\psi_X. 
\label{leadingedge3}
\end{eqnarray}

\subsection{The asymptotic expansion for $\psi_X$ in terms of  
similarity variables of the diffusion equation}

As we already pointed out above, in dominant order, the dynamical
equation (\ref{leadingedge2}) for $\psi(\xi,t)$ is a diffusion equation,
and this was reflected by the fact that  in the fully linear
spreading problem, $\psi(\xi,t)$ is just the fundamental Gaussian
similarity solution $e^{-\xi^2/(4{\mathcal D}t)}/\sqrt{t}$
--- Cf. Eq.~(\ref{phibasic}). As explained 
in \cite{evs2,storm}, the nonlinearity in (\ref{leadingedge2})
can be interpreted as a sink for the diffusive field $\psi_X$ 
to the left of the leading edge. This imposes that in contrast
to the linear problem, $\psi$ has to increase linearly in $\xi$ for
small $\xi$. The relevant fundamental solution of 
the diffusion equation which has this behavior is
\begin{equation}
\psi(\xi,t) \sim \frac{\xi}{t^{3/2}} e^{-\xi^2/(4{\mathcal D}t)},
\label{psiasymp}
\end{equation}
and as explained in detail in \cite{evs2,wimreview} one can already
obtain the dominant term of the power law relaxation of the velocity
and front shape from this argument. 

The expansion is systematized by working in the $\xi_X$ frame, as
explained above, and by recognizing that the similarity variable of
the diffusion equation is
\begin{equation}
z = \frac{\xi^2_X}{4{\mathcal D}t}.
\end{equation}
In short, since far ahead of the front in the leading edge, $\psi_X$
will fall off like a Gaussian $e^{-z}= e^{-\xi^2/(4{\mathcal D}t)}$
for a sufficiently steep front (\ref{steep}) (see \cite{evs2}), we write 
\begin{equation}
\psi (\xi_X,t) = G(z,t) e^{-z}. \label{psitoG}
\end{equation}
To ensure the Gaussian decay for large $\xi_X$ and finite $t$,
we require
\be
\label{bc}
\lim_{z\to\infty}G(z,t) \; e^{-z}=0
~~~~~~\Longleftrightarrow~~~~~~
\lim_{\xi_x \to\infty} \psi_X(\xi_X,t)=0.
\ee
Note that as we already stated in (\ref{Deq}), $\mbox{Re}\;{\mathcal D}>0$, so the
limit $z\to \infty$ should be taken along a line in the right complex $z$ plane.
This is the first boundary condition for $G$. The second boundary 
or matching condition arises from the behavior for small $\xi_X$, 
actually in the transition towards the nonlinear regime. 
In agreement with the intuitive argument about the nonlinearity
as a sink for the diffusion process, one derives
\begin{equation}
\psi_X(\xi_X,t)
\stackrel{\xi_X/\sqrt{t}\to0}{=} \alpha \xi + \beta
~~~~~~\Longleftrightarrow~~~~~~
G(z,t)=2\alpha\sqrt{{\mathcal D}zt}+\ldots,
\label{matchingcond}
\end{equation}
where $\alpha$ and $\beta$ are in general complex constants
with $\alpha\ne0$ due to the nonlinearity\footnote{For the nonlinear 
diffusion equation, we derived ${\mathcal D}\alpha=\int_{-\infty}^\infty
d\xi\;{\mathcal N}\;\psi$ in section 2.5.2 of \cite{evs2}.
The relation between non-vanishing $\alpha$ and ${\mathcal N}$
can be generalized to pattern forming fronts \cite{storm}.
In general, ${\mathcal N}$ then becomes time dependent and some temporal
averaging is required. For the cubic CGL equation (\ref{CGL}),
however, we obtain
${\mathcal D}\alpha=\int_{-\infty}^\infty {\rm d} \xi\;\; (1+ic_3) |{A}
|^2\psi(\xi)$ without temporal averaging. The phase of $\alpha$ changes 
in the same way as the phase of $\psi$ while the complete problem
is phase invariant.}.

Upon substitution of (\ref{psitoG}) into Eq.~(\ref{leadingedge3}) 
for $\psi_X$, and using the expansion (\ref{Xeq}) for $X(t)$ 
and (\ref{Gammaeq}) for $\Gamma(t)$, we obtain the equation of motion for $G$
\begin{eqnarray}   
\label{Geq}
t\partial_t G 
&-& 
 \left( c_1 + \frac{c_{3/2}} {\sqrt{t}} \right) 
 \left[ ik^* + \frac{\sqrt{z}} {\sqrt{{ \sf D} t}} 
( \partial_z -1) \right] G -
i \left( d_1 + \frac{d_{3/2}}{\sqrt{t}}  \right)  G   = 
\nonumber \\
  & &  \hspace*{0.0cm}  \left[z\partial_z^2+\left(\frac{1}{2}-z\right)
\partial_z-
    \frac{1}{2}   \right] G \nonumber \\
   & &  \hspace*{0.6cm}  +\frac{{\mathcal D}_3\sqrt{z}}{{\mathcal D}^{\frac{3}{2}} 
\sqrt{t}} \left[ \frac{3}{2}
      \left(\partial_z-1\right)^2+z \left(\partial_z-1\right)^3
    \right] G    \\
   & &  \hspace*{0.6cm}  + w\;\frac{\sqrt{z}}{\sqrt{{\mathcal D}t}}\;
\Big[t\partial_t-z(\partial_z-1)-1-ik^*c_1-id_1\Big]\;(\partial_z-1)\;G
+ \cdots .    \nonumber 
  \end{eqnarray}

The relevant long-time asymptotics of $\psi_X$ then directly follows
from solving this equation with boundary conditions (\ref{bc}) and 
(\ref{matchingcond}) \cite{evs2}. As in \cite{evs2}, the coefficients
$c_i$ and $d_i$ in $X(t)$ and $\Gamma(t)$ can be obtained by
expanding $G(z,t)$ as an asymptotic series in terms of functions of
the similarity variable $z$,
\begin{equation}
G(z,t)= t^{1/2}  g_{-1/2} (z) + g_0(z) + \frac{g_{1/2} (z)}{ t^{1/2}} +
\frac{g_{1}(z)}{t} \cdots, \hspace*{1cm} (t\gg 1), \label{Gexpansion} 
\end{equation}
where the matching condition (\ref{matchingcond}) implies that the
leading order indeed is $\sqrt{t}$ with the coefficient 
$g_{-1/2}(z) = \sqrt{z}+\ldots$ for  small $z$. 

From here on, the analysis is just the technical implication of the
expansion introduced above. Since the structure of the analysis
follows essentially the one given in our earlier work on uniformly
translating fronts, we relegate the details to appendix
\ref{appendix1}. The final outcome of the analysis is that the velocity relaxes to $v^*$
according to the general formula
\begin{equation}
v(t) \equiv v^*+\dot{X}(t) = v^* - \frac{3}{2\lambda^*t} +\, \frac{3 \sqrt{\pi}}{2
(\lambda^*)^2 t^{3/2} }\, {\rm Re}\frac{1}{\sqrt{\mathcal D}} + {\mathcal O}\left( \frac{1}{t^2}\right) , \label{v(t)2relaxation}
\end{equation}
while the phase relaxation is governed by a similar expression,
\begin{equation}
\dot{\Gamma}(t)  = -q^* \dot{X}(t) 
- \, \frac{3 \sqrt{\pi}}{2
\lambda^*  t^{3/2} }\, {\rm Im}\frac{1}{\sqrt{\mathcal D}} + {\mathcal
O}\left( \frac{1}{t^2}\right) .
\label{gamma(t)}
\end{equation}

\subsection{Convergence of a coherent front profile to its asymptotic shape}

The above expressions are valid for any pulled front, irrespective 
of whether it is asymptotically uniformly translating or a coherent 
or incoherent pattern forming front\footnote{
Of course, for uniformly translating fronts there is no phase, 
hence $q^*=0=\Omega^*$ and Im$~{\mathcal D}=0$ in (\ref{gamma(t)}).}.
Here `coherent' means that the approximately periodic pattern laid down by
the leading edge of the front stays periodic in the nonlinear region,
while incoherent means that the pattern undergoes some further
dynamics behind the front. Such 
incoherent fronts arise e.g. in some parameter regimes
of the cubic and quintic Complex Ginzburg Landau equation
\cite{wimreview,storm,nb,vsh} or the Kuramoto-Sivashinsky equation
\cite{wimreview}. Even when a pulled pattern forming front is incoherent
the linear dynamics in the leading edge  is
described by the above equations. The dynamics in the leading edge is
therefore still coherent: the incoherent behavior only sets in in the
region where the dynamics become truly nonlinear. Since the matching
condition which the nonlinear dynamics imposes on  the linear leading
edge dynamics is still the same in this case
\cite{wimreview,storm}, the above results even apply to incoherent
fronts. However, the phase relaxation applies in that case only to the
coherent dynamics in the leading edge. 

If the pattern forming front is coherent, the results apply 
throughout the whole front region. More precisely, we call a front 
coherent if the asymptotic front
solution is time periodic in the co-moving frame $\xi=x-v^*t$, i.e. if
there is some period $T$ such that
\begin{equation}
\Phi(\xi,t+T) = \Phi(\xi,t),\hspace*{1cm}
\mbox{where }~\phi(\xi,t) \stackrel{t\to \infty}{=} \Phi(\xi,t) .
\end{equation}
The dynamics of the leading edge actually determines this period to be
\be
T=2\pi/\Omega^*,
\ee
where $\Omega^*$ is the frequency determined by the saddle point 
(\ref{saddlepoint}).
This can be easily read from Eq.~(\ref{phibasic}) or from 
Eq.~(\ref{leadingedgetrafo2}) and the knowledge that 
$\psi_X(\xi_X,t)$ becomes stationary for $t\to\infty$.

Because of the temporal periodicity, we can generally write a
coherent $\Phi$ in the whole spatial domain as a Fourier series 
\begin{equation}
\Phi(\xi,t) = \sum_{n=0,\pm 1,\ldots} e^{-in\Omega^*  t} \Phi^n(\xi).
\end{equation}

In our analysis \cite{evs2} of fronts which converge to a uniformly 
translating front solution, we  explicitly showed that to order ${\mathcal
O}(1/t^2)$, the front shape relaxation follows the velocity relaxation
adiabatically. 
An extension of the analysis  to coherent pattern
forming fronts shows that a similar result holds for these. The reason is that when the front  is converging to
its asymptotic shape  as $1/t$, the temporal derivative terms in the dynamical equations
only generate terms of order $1/t^2$ in the asymptotic expansion, while the terms coming
from the adiabatic variation of $v(t)$ and $\Gamma(t)$ generate terms of order $1/t$ and $1/t^{3/2}$.
In other words, to order $1/t^{3/2}$ the only temporal dependence comes in parametrically via
$v(t)$ and $\Gamma(t)$. Thus, for long
times, {\em coherent pattern forming} fronts relax to their asymptotic shape
according to
\be
\phi(x,t)  \stackrel{t\gg1} {=}   \Phi_{v(t)} ( \xi_X,t) + {\mathcal O} (t^{-2})~~\mbox{with} ~~\Phi_(\xi_X,t) \approx \Phi_{v(t)}(\xi_X,t+T(t)), 
\ee
where $v(t)$ and $\Gamma(t)$ are given by Eqs~(\ref{v(t)2relaxation})
and (\ref{gamma(t)}) above, and where $T(t)$ is the instantaneous
period $2\pi / (\Omega^*+\dot{\Gamma}(t) )$.  In terms of the temporal Fourier series
this result can be written as
\be
\phi(x,t)  \stackrel{t\gg1} {=}   \sum_{n=0,\pm 1,\cdots} 
e^{-in(\Omega^*t+ \Gamma(t)) }
\Phi_{v(t)}^n(\xi_X)  + {\mathcal O}(t^{-2}) \label{shaperelaxation}
\ee
where the $\Phi^n_{v}$ are the Fourier
transform functions of the coherent pattern forming
solutions\footnote{Clearly, this result implies the existence of a
two-parameter family of coherent front solutions, parametrized by
their velocity and frequency. It is argued in \cite{wimreview} that this is
the generic case, and that if such a two-parameter family of solutions
does not exist, there generically does not exist a coherent pulled
front solution either; the fronts will then be incoherent.}
with velocity $v$ and frequency $\Omega^*+\dot{\Gamma}$. Thus the
above result expresses that the coherent front profiles follow  this family of
solutions adiabatically, and that their velocity and frequency shift
$\dot{\Gamma}$ is set completely by the dynamics in the leading edge.

\section{Numerical study of the relaxation behavior of fronts in the 
Swift-Hohenberg equation}

We now illustrate the above analysis with numerical results obtained
for the Swift-Hohenberg equation (\ref{swifthohenberg}). 
This equation has often been used
\cite{dee,vs2,collet1,collet2,eckmann,collet3}
as one of the simplest equations to illustrate the behavior of
coherent pattern forming fronts. Collet and Eckmann were the first to
prove that fronts propagating into the linearly unstable state $\phi=0$
are pulled; the analysis of section 2 applies too and therefore
establishes this fact as well. In the simulations of this equation
presented here, we study the approach of the fronts to these
asymptotic pulled front solutions, starting from a Gaussian initial
condition. Note in this regard that while the
Swift-Hohenberg is often studied for small $\varepsilon$ where the
dynamics maps onto an amplitude expansion, our front convergence
analysis applies generally. We will illustrate this by taking finite
values of $\varepsilon$. Fig.~\ref{figprofile} shows a $\phi$-profile
for $\varepsilon=0.5$.

We first illustrate an important ingredient of our convergence
analysis. As we argued above,  in the
co-moving frame $\xi=x-v^*t$  the leading edge
variable $\psi$ defined in (\ref{leadingedgetrafo}) should
asymptotically behave as $\xi/(t^{3/2})e^{-\xi^2/(4{\mathcal
D}t)}$ [Cf. Eq.~(\ref{psiasymp})]. To illustrate  this for the
Swift-Hohenberg equation, we show 
in Fig.~{\ref{fig1} three snapshots of the leading edge variable
$t^{3/2} \hat{\psi}(x,t) = e^{\lambda^*(x-v^*t)}\;\phi(x,t)$ in a
simulation for $\varepsilon=0.5$; according to our analysis, the
envelope of this function should asymptotically behave as  
\begin{equation}
(x-v^*t)\; e^{-(x-v^*t)^2/(4Dt)}, \hspace*{1cm} \mbox{with}~
\frac{1}{D} \equiv {\rm Re}\,\frac{1}{\mathcal D}\label{shasymp}.
\end{equation}
Our numerical results in Fig.~\ref{fig1} fully confirm this behavior.

\begin{figure}[t]
\begin{center}
\epsfig{figure=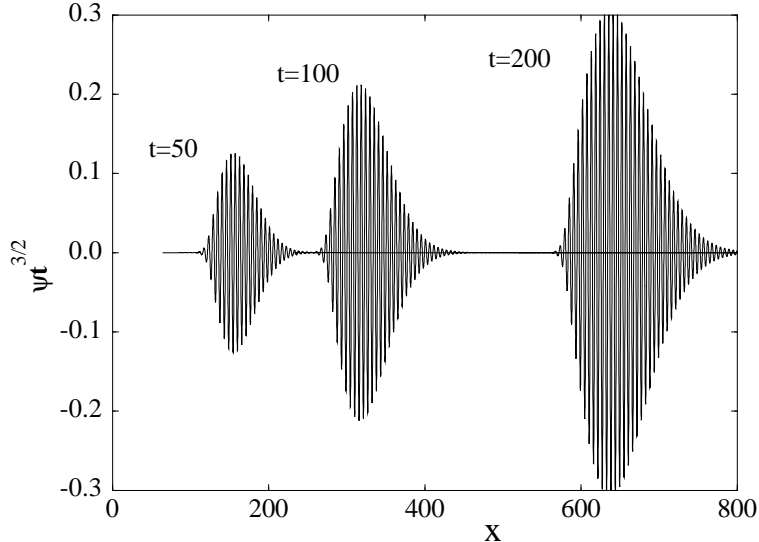,angle=-90,width=0.7\linewidth}
\end{center}
\caption[]{Three snapshots of the function $t^{3/2} \hat\psi$ obtained 
from our simulations
of the Swift-Hohenberg equation for $\varepsilon=0.5$. The results
confirm the asymptotic behavior (\ref{shasymp}). Note in particular
the diffusive broadening of the pattern: the one at time $t=200$ is twice
as wide as the one at time $t=50$. }\label{fig1}
\end{figure}

To test our convergence results, we have to
extract the velocity $v(t)$ and frequency $\Omega^*+ \dot{\Gamma}(t)$
from our numerical data. Because of the oscillating character of the
fronts, this is nontrivial in principle.
We will do it in a pragmatic way, replacing differentials by finite
difference approximants: In our simulation, we keep track of the local
maxima of $\phi(x,t)$ and from these  determine the
positions $X_n$ and times $t_n$ at which the foremost maximum $n$ reaches a
predetermined fixed ``level'' $\ell$.  From this we calculate the
finite difference approximants
\begin{equation}
\label{difference}
v_{\ell}(t_n) = \frac{X_n- X_{n-1}}{t_n-t_{n-1}}, \hspace*{1cm} 
\Omega_{\ell}(t_n) = \frac{2\pi }{t_n-t_{n-1}} .
\end{equation}
and then analyze whether indeed the convergence of these quantities to their
asymptotic values is consistent with the  universal $\ell$-independent
behavior derived above. The error of the finite difference approximants 
is of ${\mathcal O}(1/t^2)$ only. For testing the convergence up 
to terms of ${\mathcal O}(1/t^{3/2})$, the discretization error 
is therefore irrelevant.

In Fig.~\ref{figv} we show two plots of the velocity 
relaxation data for two different values of $\varepsilon$, namely
$\varepsilon=0.5$ and $\varepsilon=5$. The various  lines
indicate the velocity extracted for different levels $\ell$. To probe
the predicted behavior in detail, we have plotted $v_\ell(t)-v^*-c_1/t$
versus $t^{-3/2}$. According to our prediction (\ref{v(t)2relaxation})
this velocity difference should asymptotically approach 0 along the
dashed lines. Similar plots for the frequency relaxation, obtained
from the same runs, are shown in Fig.~\ref{figomega}. 
Clearly, all our numerical results are in full agreement
with the predicted behavior.

We finally study the convergence of the shape of the profile to its
asymptotic form. In principle, the information is contained in the
expression (\ref{shaperelaxation}) above, but to make it explicit 
one would have to know
all functions $\Phi_v^n$. Since our goal here is simply to check that
the shape relaxation follows the velocity and phase relaxation
adiabatically, we circumvent this problem as follows. We construct an
effective (real) envelope $A(\xi_X,t) $ of the front profile\footnote{Note 
that this real envelope $A$ differs from the complex amplitude $A$ 
of the previous sections.}  
in the co-moving frame  by tracking the positions of the maxima of
$\phi(x,t)$ during one effective period $2\pi/(\Omega^*+
\dot{\Gamma}(t))$. In doing so, $\xi_X$ is determined by requiring that 
$A(\xi_X=0,t)= const.$ where the constant is chosen so that the
level of the
effective envelope at this point is about half of its asymptotic
value. The implication of (\ref{shaperelaxation}) now is that the
convergence of the 
effective envelope $A(\xi,t)$ determined this way should, up to
terms of ${\mathcal O}(1/t^{2})$, adiabatically follow the velocity
and shape relaxation:
\begin{equation}
A(\xi_X,t) = A_{v(t),\dot{\Gamma}(t)}(\xi_X) + {\mathcal O}(1/t^2),
  \end{equation}
so that 
\begin{eqnarray}  A(\xi_X,t ) - A(\xi_X,t^\prime)   & = & \frac{\delta
    A_{v,\dot{\Gamma}}(\xi_X)}{\delta v} \, [v(t)-v(t^\prime)]
  \nonumber \\ & & +
  \frac{\delta A_{v,\dot{\Gamma}} (\xi_X) }{\delta \dot{\Gamma}}
\,  [\dot{\Gamma}(t)-\dot{\Gamma}(t^\prime)] + {\mathcal O}(1/t^2).\label{Aeq}
\end{eqnarray}
As in the discretization (\ref{difference}), the averaging over one period 
only affects the terms of ${\mathcal O}(1/t^2)$ in this expression.

\begin{figure}[t]
\begin{center}
\hspace*{-4mm}
{\tt (a)} 
\hspace*{-2mm} 
\epsfig{figure=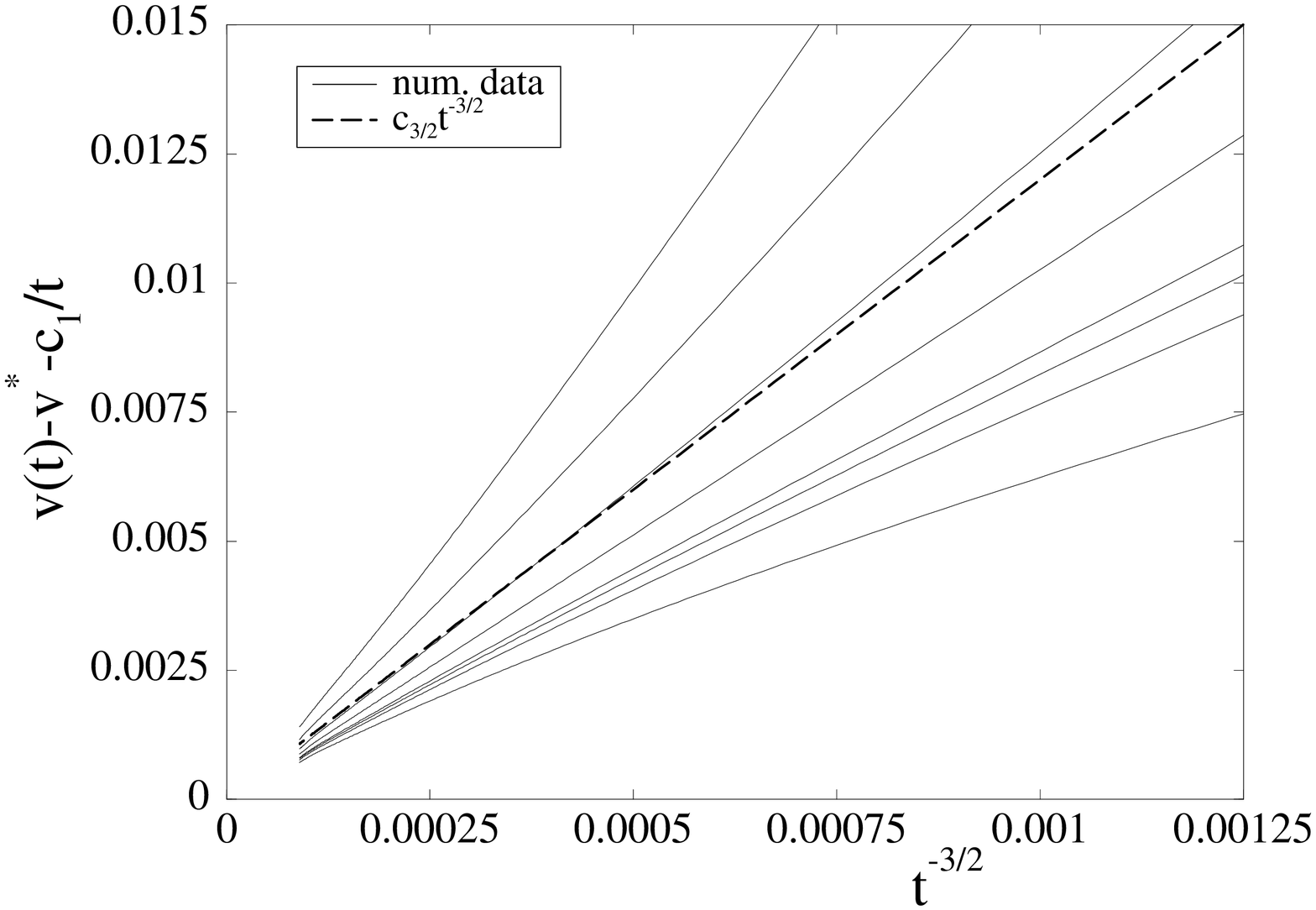,angle=-90,width=0.42\linewidth}
\hspace*{0.1cm}
{\tt (b)} \hspace*{-1mm}
\epsfig{figure=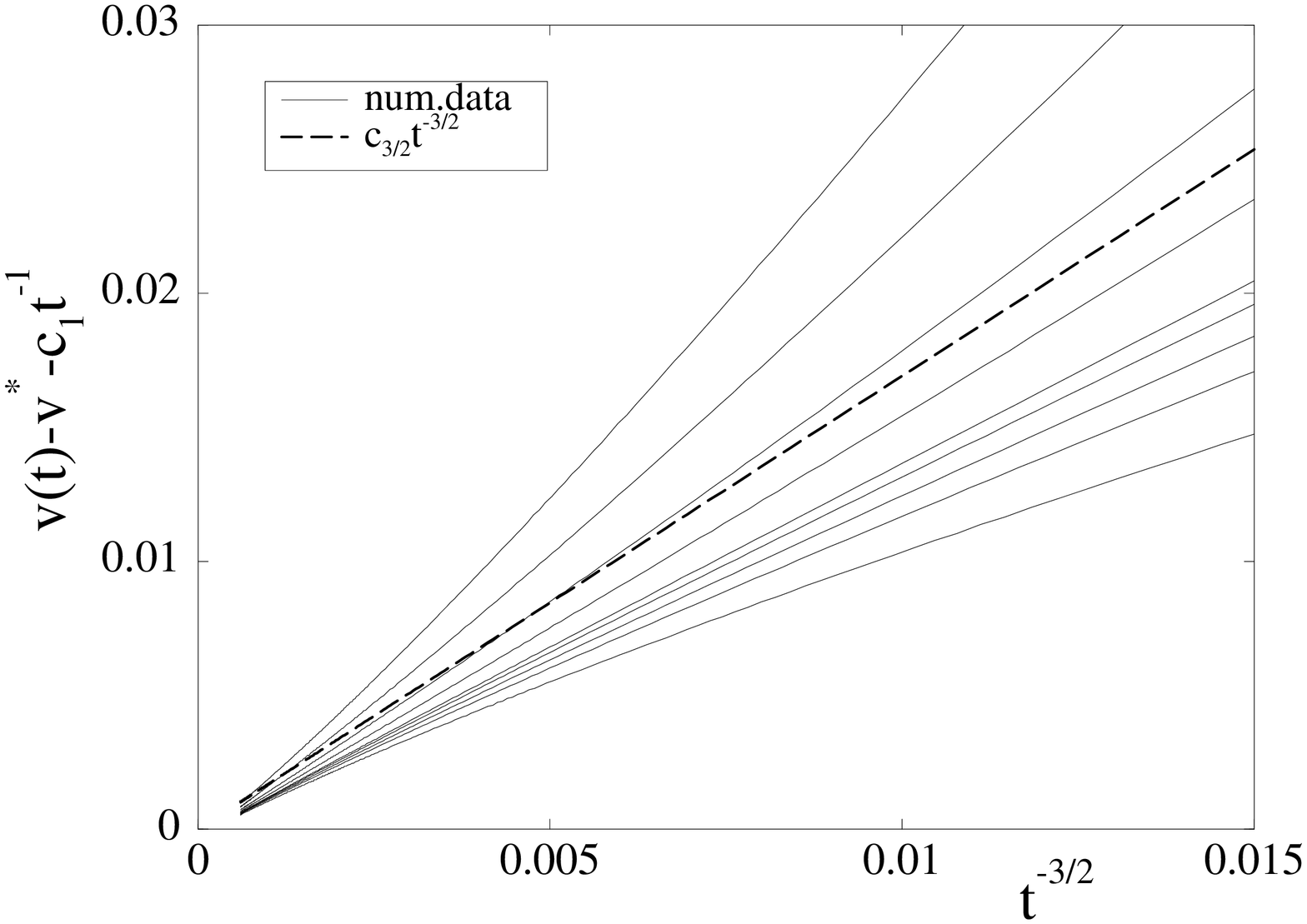,angle=-90,width=0.42\linewidth}
\end{center}
\caption[]{Velocity difference $v_{\ell} (t) - v^*-c_1/t$ as a
function of $t^{-3/2}$ for $\varepsilon=0.5 $ (panel a) and
$\varepsilon=5$ (panel b). The various lines denote, from top to
bottom, the levels $\ell=0.0001 \sqrt{\varepsilon}$,
$0.001\sqrt{\varepsilon}$, $0.01\sqrt{\varepsilon}$,
$0.05\sqrt{\varepsilon}$, $0.2\sqrt{\varepsilon}$,
$0.3\sqrt{\varepsilon}$ and $0.5\sqrt{\varepsilon}$. The dashed line is
the asymptotic slope according to the exact expression
(\ref{v(t)2relaxation}). }\label{figv}
\end{figure}

\begin{figure}[t]
\begin{center}
\hspace*{-4mm}
{\tt (a)} 
\hspace*{-2mm} 
\epsfig{figure=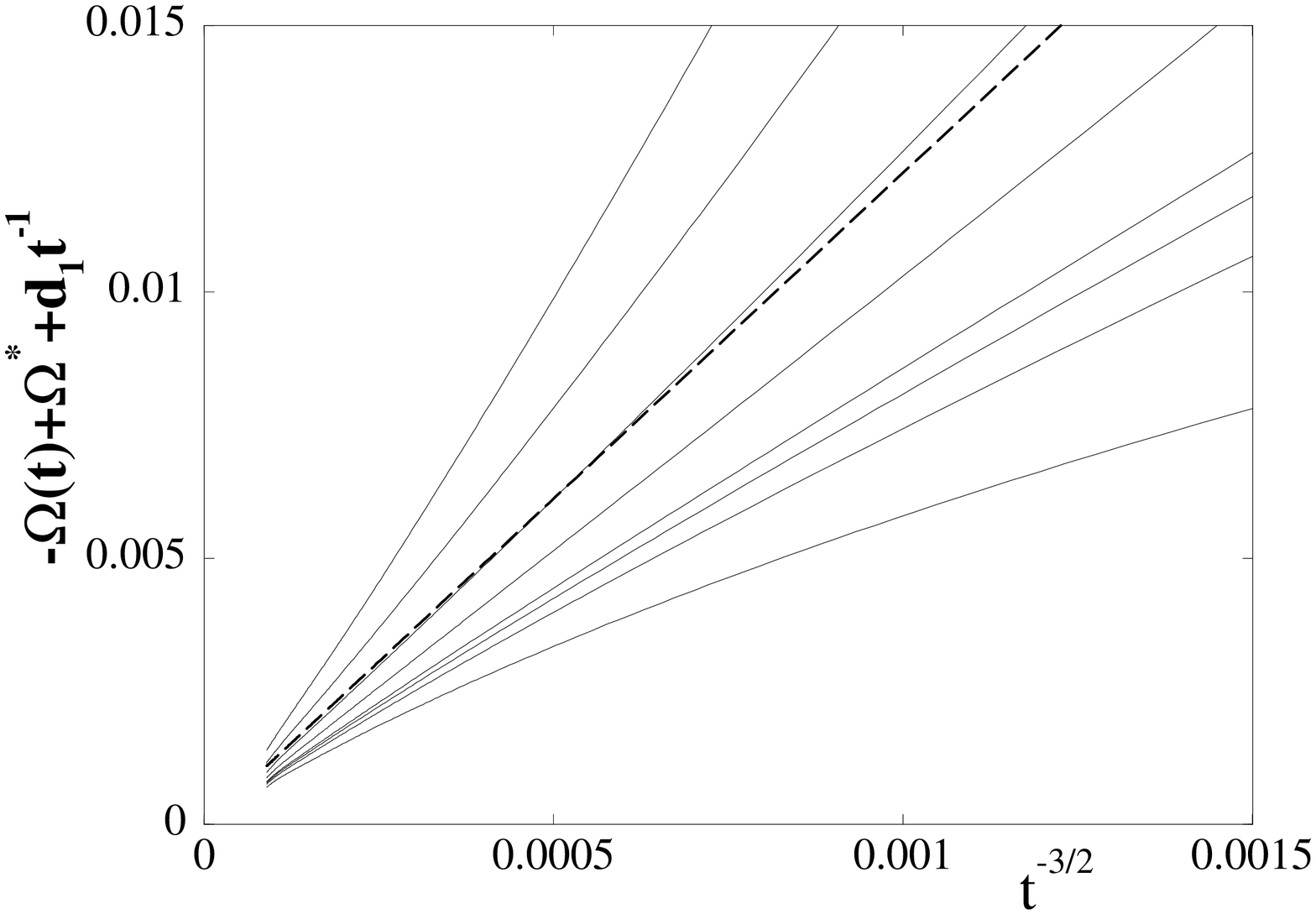,angle=-90,width=0.42\linewidth}
\hspace*{0.1cm}
{\tt (b)} \hspace*{-1mm}
\epsfig{figure=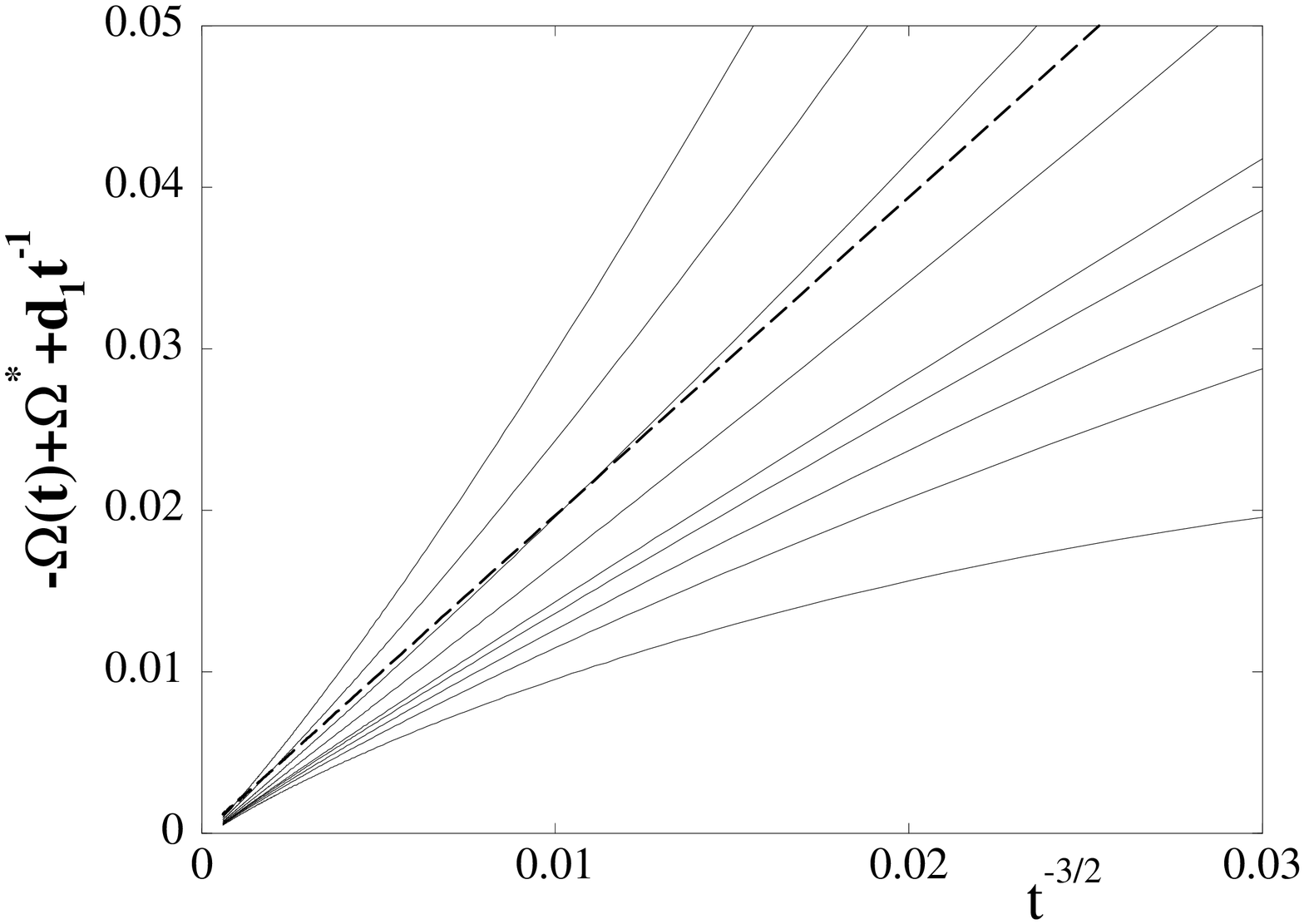,angle=-90,width=0.42\linewidth}
\end{center}
\caption[]{As Fig.~\ref{figv}, but now for the frequency relaxation
$\Omega(t)=\Omega^*+\dot{\Gamma}(t)$.}\label{figomega}
\end{figure}

\begin{figure}[t]
\begin{center}
\epsfig{figure=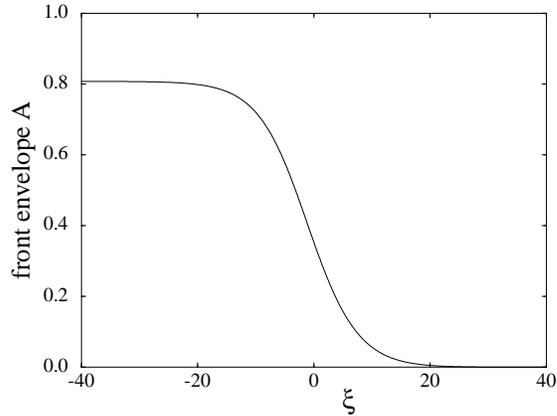,angle=-90,width=0.50\linewidth}
\end{center}
\caption[]{The front envelope $A(\xi_X,t)$ for $\varepsilon=0.5$
obtained as described in the text. In this case, $t=160$, and the
front shape is obtained by averaging over one period that lasts about 
$\Delta t = 2$. Note the different horizontal scale
in comparison with Fig.~\ref{figprofile}.} \label{figshape}
\end{figure}

Fig.~\ref{figshape} shows the effective envelope $A(\xi_X,t)$ for
the front from Fig.~\ref{figprofile}. The
figure confirms that even for this value, where the pattern behind the 
front is rapidly oscillating, the effective envelope can be obtained
accurately and is smooth.

In Fig.~\ref{figshaperel} we present our analysis of the large-time shape
relaxation of this profile. Panel (a) shows the difference
$A(\xi_X,t)-A(\xi_X,180)$, while in panel (b) we plot the ratio
\begin{equation}
\frac{A(\xi_X,t)-A(\xi_X,180)}{1/t - 1/180},\label{Aratio}
\end{equation}
which according to our prediction (\ref{Aeq}) {\em should for large
  times become  a function of $\xi_X$ only}. It is clear that 
our numerical results fully corroborate this.

\begin{figure}[t]
\begin{center}
\hspace*{-4mm}
{\tt (a)} 
\hspace*{-2mm} 
\epsfig{figure=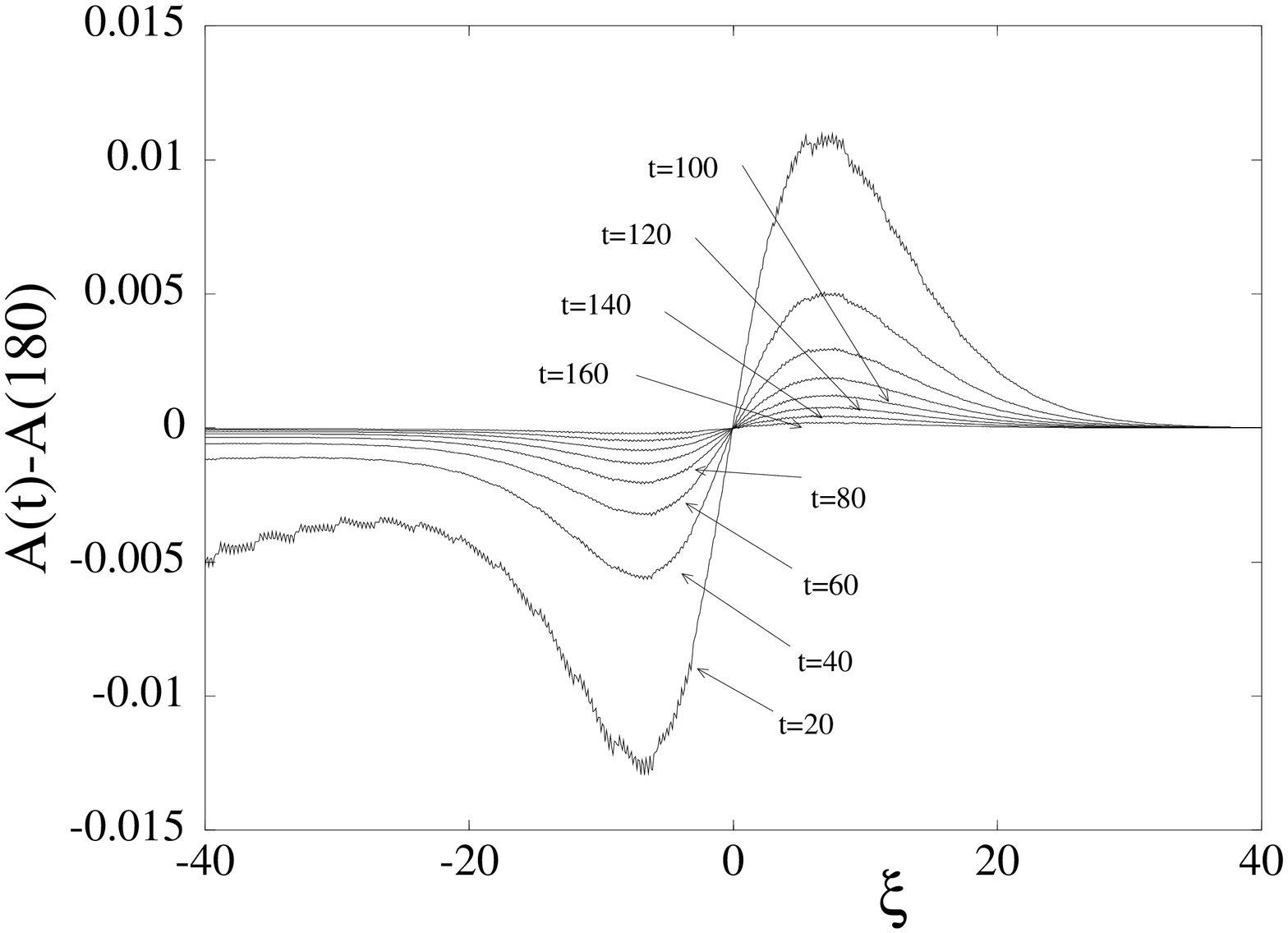,angle=-90,width=0.44\linewidth}
\hspace*{0.1cm}
{\tt (b)} \hspace*{-1mm}
\epsfig{figure=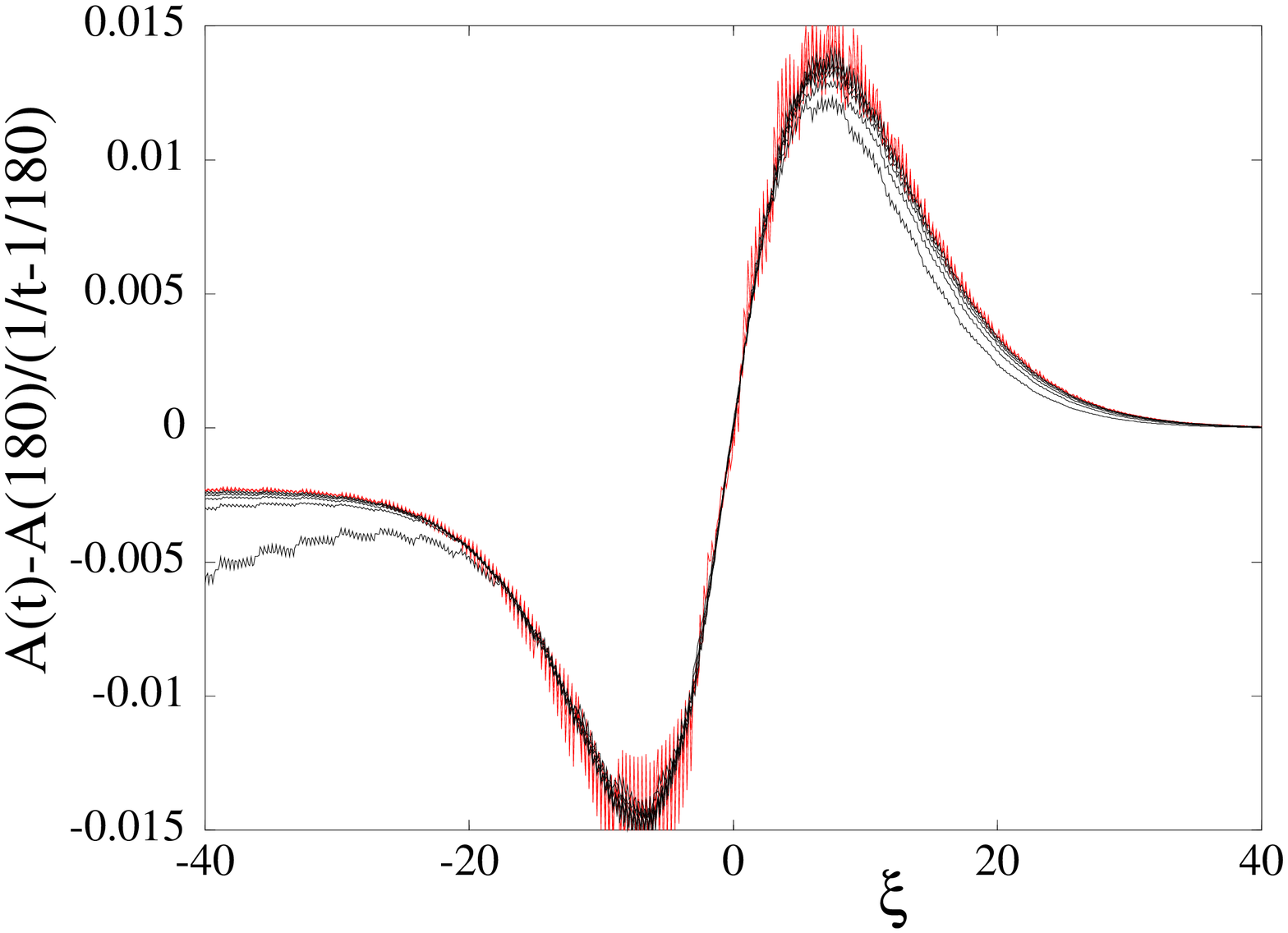,angle=-90,width=0.44\linewidth}
\end{center}
\caption[]{(a) The convergence of the effective envelope difference
  $A(\xi_X,t)-A(\xi_X,180)$, as obtained from the numerical solutions
  illustrated in Fig.~\ref{figshape}.  (b) The ratio (\ref{Aratio})
  as obtained from the data shown in panel (a). The figure confirms
  that this ratio converges to a time-independent function, in
  agreement with our predictions.  }\label{figshaperel}
\end{figure}

\newpage

\section{Conclusion}

In this paper we have presented two types of results. First of all, we
have introduced a simple line of analysis which allows us to prove for 
certain classes of equations which include the Swift-Hohenberg
equation, the Extended Fisher-Kolmogorov equation and the cubic
Complex Ginzburg Landau equation that fronts are
pulled. The method works for real or complex equations 
and fields and is not restricted to nonlinearities like 
${\mathcal N}(A)\;A=|A|^{2n}A$ with integer $n$,
but also can treat nonlinearities that depend, e.g., on $\partial_xA$.
Important is that the over-all sign of Re~${\mathcal N}$ can be determined.

Second, we have derived the universal slow convergence of
the velocity and phase of coherent pattern forming pulled fronts to
their asymptotic value. Numerical simulations of the Swift-Hohenberg
equation are in full agreement with these predictions. In another
paper \cite{storm}, we have shown that the results for the velocity
convergence also apply to incoherent pattern forming fronts.

We are grateful to Kees Storm for many useful discussions. 

\newpage

\begin{appendix}
\section{Derivation of Eqs.~(\ref{v(t)2relaxation}) and (\ref{gamma(t)})}
\label{appendix1}

The derivation follows essentially the lines of \cite{evs2},
except that $z$ is now a complex rather than a real variable,
and that there are additional terms due to $q^*$ and $\dot\Gamma$.
The task is to solve (\ref{Geq}) with the ansatz (\ref{Gexpansion})
and with boundary conditions (\ref{bc}) and (\ref{matchingcond}). 
Actually, the analysis of the nonlinear region for finite $t$
contributes additional terms to (\ref{matchingcond}) which will
play a role in the calculation of the subleading terms.
The boundary conditions for $\psi_X$ become
\ba
&\psi_X(\xi_X,t)&=\alpha\xi_X+\beta+\frac{f_1(\xi_X)}{t}
+O\left(\frac{f_{3/2}(\xi_X)}{t^{3/2}}\right),
\\
&\psi_X(\xi_X,t)&\stackrel{\xi_X^2/(4{\mathcal D}t)\gg1}{\longrightarrow}0.
\ea
Insertion into the ansatz (\ref{Gexpansion}) implies for the function
$G(z,t)$ that
\ba
\label{bcg1}
&&G(z,t)=\sqrt{t}\Big[2\alpha\sqrt{{\mathcal D}z}+O\left(z^{3/2}\right)\Big]
+\Big[\beta+O\left(z\right)\Big]+\frac{O\left(\sqrt{z}\right)}{\sqrt{t}}
+O\left(\frac1{t}\right)~,\nonumber\\&&\\
\label{bcg2}
&&\lim_{z\to\infty} e^{-z}\;G(z,t)=0.
\ea
These boundary conditions determine a unique solution for the functions
$g_{1/2}(z)$ and $g_0(z)$ and the constants $c_1$, $d_1$, $c_{3/2}$
and $d_{3/2}$ in $\dot X$ and $\dot\Gamma$, as we will derive below.

Inserting (\ref{Gexpansion}) into (\ref{Geq}), we see that the dominant 
terms are of order $t^{1/2}$. Upon collecting these, we get
\begin{equation}
     \left[ z\frac{d^2}{dz^2} + \left(\frac{1}{2}-z\right)\frac{d}{dz}
      -1-\lambda^* c_1+i(d_1+q^* c_1) \right] g_{-1/2}=0. 
\label{g-1/2eq}
\end{equation}
This homogeneous equation is an example of Kummer's equation \cite{abramowitz}
\begin{equation}
  \label{eq:Kummer}
\hat{T}[a,b] g \equiv  \left[z\frac{d^2}{dz^2}+(b-z)\frac{d}{dz}-a\right]g=0,
\end{equation}
whose general solution is a superposition of the two confluent 
hypergeometric functions
\begin{equation}
M(a,b,z)~\mbox{ and }~z^{1-b}M(1+a-b,2-b,z). \label{Meqs}
\end{equation}
These functions are defined through the series
\begin{equation}
  M(a,b,z)=\sum_{n=0}^\infty \frac{(a)_nz^n}{(b)_nn!}, \label{kummerM}
\end{equation}
where
\begin{equation}
  (a)_n=a(a+1) \ldots (a+n-1) = \frac{\Gamma(a+n)}{\Gamma(a)}, ~~ ~~~(a)_0=1. 
\end{equation} 
The asymptotic large-$z$ behavior of the functions $M$ for positive
$b$ is
\begin{equation} \label{3050} 
  M(a,b,z) \stackrel{z\to\infty}{\simeq}\left\{
\begin{array}{ll}\frac{\Gamma(b)}{\Gamma(a)}\; z^{a-b}\;e^z 
  & \mbox{ for } a \neq 0,-1,-2,-3,\cdots,\\ 
  \frac{(a)_{|a|}z^{|a|}}{(b)_{|a|} {(|a|)}!} & \mbox{ for }
  a = 0, -1, -2, -3,\cdots,
\end{array}\right.\end{equation} 

Let us return to Eq.~(\ref{g-1/2eq}) for $g_{-1/2}(z)$. 
The boundary condition (\ref{bcg1}) implies
\be
\label{bcg11}
g_{-1/2}(z)=2\alpha\sqrt{{\mathcal D}z}+O(z^{3/2}).
\ee
Since $M(a,b,z=0)=1$, a contribution of the solution $M(a,b,z)$ is excluded
through (\ref{bcg11}), and $g_{-1/2}(z)$ has to be proportional
to $z^{1-b}M(1+a-b,2-b,z)$. With boundary condition (\ref{bcg1}),
we therefore get 
\begin{equation}
g_{-1/2} (z) = 2\alpha\sqrt{{\mathcal D}z} \;
M \left(\frac{3}{2}+\lambda^* c_1-i(d_1+q^* c_1 ),\frac{3}{2},z\right).
\end{equation}

Furthermore, (\ref{3050}) shows that the Kummer functions $M(a,b,z)$
diverge as $e^z$ when the coefficient $a$ is not zero or a negative
integer,  while they are simple polynomials when $a$ is zero or a 
negative integer since then the coefficients $(a)_n$ vanish for $n\ge
1-a$.   An exponential divergence of $g$ is not allowed according to 
the second boundary condition (\ref{bcg2}); this fixes
\begin{equation}
1+a-b=\frac{3}{2}+ \lambda^*c_1- i(d_1+ q^* c_1)=0,-1,-2,\ldots~. 
\label{cdeq}
\end{equation}
For a detailed discussion of the solutions with $1+a-b=-1,-2,\ldots$,
we refer to \cite{evs2}: essentially, these solutions are dynamically
not relevant since they will always be overrun by the solution with $1+a-b=0$.
As both $c_1$ and $d_1$ are real, (\ref{cdeq}) with $1+a-b=0$ implies
\begin{equation} 
  c_1=-\frac{3}{2\lambda^*}, \hspace*{1cm}  d_1=- q^* c_1~, 
\label{c1d1eq}
\end{equation}
with the corresponding solution
\begin{equation}
  g_{-1/2} (z)= 2 \alpha \sqrt{{\mathcal D} z}. \label{g-1/2sol}
\end{equation}

The terms of order $t^0$ obtained by  subsituting (\ref{Gexpansion})
into (\ref{Geq}) are
\begin{eqnarray}
\label{g0eq}
 & & \hat{T}\left[\half +\lambda^* c_1- i(d_1+q^* c_1),\half
\right]   g_0(z) =
\nonumber \\
 & &\hspace*{0.8cm}  \left[ - ik^*c_{3/2} -  c_1 \frac{\sqrt{z}}{\sqrt{\mathcal
    D}}\left(\partial_z-1\right) -i d_{3/2} \right] g_{-1/2}(z)
\nonumber \\  & &\hspace*{1.6cm}
- \frac{{\mathcal D}_3\sqrt{z}}{{\mathcal D}^{\frac{3}{2}}  } \left[ \frac{3}{2}
      \left(\partial_z-1\right)^2+z \left(\partial_z-1\right)^3
    \right] g_{-1/2}(z) \\
   & &  \hspace*{1.6cm}  - w\;\frac{\sqrt{z}}{\sqrt{\mathcal D}}\;
\left[\frac1{2}-z(\partial_z-1)-1-ik^*c_1-id_1\right]\;(\partial_z-1)\;
g_{-1/2}(z).
\nonumber
\end{eqnarray}
The function $g_{-1/2}(z)$ on the right hand side of (\ref{g0eq}) is
known from (\ref{g-1/2sol}); likewise $c_1$ and $d_1$ are known from
(\ref{c1d1eq}). Substitution of these results gives the 
following inhomogeneous equation for $g_0(z)$
\begin{eqnarray}
 \hat{T}[-1, \half]  \,  g_{0}(z) &=&
  2 \alpha \left[ c_{3/2}\lambda^* -i(d_{3/2} +q^* c_{3/2} )  \right] 
\sqrt{{\mathcal D}z} + \frac{3\alpha}{2\lambda^*} 
\left(1-2z\right) 
\nonumber \\  & &\hspace*{0.3cm}
+2 \alpha \;\frac{{\mathcal D}_3 }{{\mathcal D}} \left[ z^2-3z+\frac{3}{4}
    \right]
+2\alpha\;w \left[z^2-3z+\frac{3}{4}\right].
\label{g0eq2}
\end{eqnarray}
The general solution of this inhomogeneous equation is a particular
solution plus the sum of two independent solutions of the homogeneous
equation $\hat{T} [-1,\half ]g_0(z)=0$. The latter can again be written 
in terms
of Kummer functions. It is easy to find particular solutions which
reproduce  most of the terms on the right by
noting that
\begin{eqnarray}
\hat{T} [-1,\half ] \sqrt{z}& =& \half \sqrt{z}, \hspace*{2cm}
\hat{T}[-1,\half ]  1 =1,\nonumber \\
\hspace*{1cm} \hat{T}[-1,\half ]  z & = &\half , \hspace*{2.4cm}
\hat{T}[-1,\half ]  z^2= -z^2+3z.
\label{Teqs}
\end{eqnarray}
With these terms, we can generate all the terms on the right hand side
of (\ref{g0eq}), except for the term linear in $z$. We can generate
this term by noting that the function
\begin{equation} \label{3058} 
  F_N(z)=\sum_{n=N}^\infty \frac{(1)_{n-2}\;
    z^n}{\left(\frac{1}{2}\right)_n\;n!}
\end{equation} 
is proportional to a truncated Kummer series $M(1,\half,z )$ (see below)
and solves
\begin{equation} \label{3059} 
 \hat{T}[-1,\half]    
F_N(z) = \frac{z^{N-1}}{\left(\frac{1}{2}\right)_{N-1}\;(N-1)},
\quad\mbox{ hence }~  \hat{T}[-1,\half] F_2(z) = 2z~.
\label{Feqs}
\end{equation} 
Using all the results (\ref{Meqs}), (\ref{Teqs}) and  (\ref{Feqs}), 
we can write the general solution of (\ref{g0eq2}) as
\begin{eqnarray}
\label{g0result}
g_0(z) &=&  k_0 (1-2z)  + l_0 \sqrt{z} M\left(-\frac{1}{2},
\frac{3}{2},z\right)  \nonumber \\
 & & \hspace*{0.8cm} + 4\alpha \left[ c_{3/2}\lambda^* -i(d_{3/2} +q^* c_{3/2}
) \right] \sqrt{{\mathcal D} z}\\ & &\hspace*{0.8cm}  +
\frac{3\alpha}{2\lambda^*}\Big[1-F_2(z)\Big]  
-2\alpha\left(\frac{{\mathcal D}_3}{{\mathcal D}}+w\right) \left[ z^2-\frac{3}{4}\right].
\nonumber 
\end{eqnarray}
where we used the fact that $M(-1,\half,z)=1-2z$. The parameters
$k_0$, $l_0$, $c_{3/2}$ and $d_{3/2}$ are again determined by the 
boundary conditions. First, the boundary condition (\ref{bcg1})
implies for $g_0$ that $g_0(z)= \beta + {\mathcal O}(z)$. This gives 
with (\ref{g0result}) 
\begin{eqnarray}
\beta + {\mathcal O}(z) &=&  \left[  k_0 + \frac{3\alpha}{2}
\left(\frac1{\lambda^*} + \frac{{\mathcal D}_3}{\mathcal D}+w\right)\right] \nonumber \\
& & \hspace*{0.4cm} 
+  \left[ 4\alpha \left( c_{3/2}\lambda^* -i(d_{3/2} +q^* c_{3/2}
) \right) \sqrt{{\mathcal D} }+ l_0 \right] \sqrt{z}  + \cdots.
\end{eqnarray}
The first term on the right determines the coefficient $k_0$ in
terms of $\alpha$, $\beta$ and the other parameters, but this term is
not needed in the sequel. The condition that the prefactor of the
$\sqrt{z}$ term on the right vanishes gives 
\begin{equation}
\left( c_{3/2}\lambda^* -i(d_{3/2} +q^* c_{3/2}
) \right) \sqrt{{\mathcal D} }+ \frac{l_0}{4\alpha} =0. \label{firsteq}
\end{equation}
Second, the boundary condition (\ref{bcg2}) imposes also for $g_0(z)$,
that the function does
not diverge exponentially for large $z$. There are two terms in
(\ref{g0result}) which diverge exponentially: the Kummer function $M$,
whose asymptotic behavior is given in (\ref{3050}), and the function
$F_2(z)$. It is easy to see that for large $z$, we have
\begin{equation}
z^2 \frac{ {\rm d}^2F_2(z) }{{\rm d}z^2} \simeq  M(1,\half,z)
~~~\Longrightarrow ~~~ F_2(z) \simeq
\sqrt{\pi} z^{-3/2} e^z~.
\end{equation}
Therefore the requirement that the two exponentially divergent
terms in $g_0(z)$ cancel each other, translates into
\begin{equation}
\frac{l_0}{4\alpha} +  \frac{3\sqrt{\pi}}{2\lambda^*}=0.\label{secondeq}
\end{equation}
Upon eliminating $l_0/\alpha$ from equations (\ref{firsteq}) and
(\ref{secondeq}) we simply get
\begin{equation}
c_{3/2} = \frac{3\sqrt{\pi}}{2(\lambda^*)^2} {\rm Re}\, \frac{1}{\sqrt{\mathcal
D}},\hspace*{1.6cm} d_{3/2} = - \frac{3\sqrt{\pi}}{2\lambda^*} {\rm
Im}\,\frac{1}{\sqrt{\mathcal D}} - q^* c_{3/2}.
\end{equation}
The second contribution to $d_{3/2}$ is just the contribution to the
phase relaxation which is induced by the
relaxation of $v(t)$. Upon substitution of these results in the
expansions (\ref{Xeq}) for $X(t)$ and (\ref{Gammaeq}) for $\Gamma(t)$
we get the results (\ref{v(t)2relaxation}) and (\ref{gamma(t)}).

\end{appendix}


\begin{thebibliography}{99}


\bibitem{briggs}  R. J. Briggs, {\em Electron-Stream Interaction with
    Plasmas} (MIT Press, Cambridge, 1964).

\bibitem{bers} A. N. Bers, {\em Space-Time Evolution of Plasma
    Instabilities --- Absolute and Convective},  
  in: {\em Handbook of Plasma Physics},
  M. N. Rosenbluth and R. Z. Sagdeev, eds. (North-Holland, Amsterdam, 1983).

\bibitem{ll} L. D. Landau and E. M. Lifshitz, 
  {\em Electrodynamics of Continuous Media},
  vol. 8 of {\em Course of Theoretical Physics} (Pergamon, New York,
1975).

\bibitem{huerre} P. Huerre, in: {\em Propagation in Systems far from
    Equilibrium}, J. E. Wesfreid, H. R. Brand, P. Manneville,
  G. Albinet, and N. Boccara, eds (Springer, New York, 1988).

\bibitem{evs1}  U.\ Ebert and W.\ van Saarloos, {\em Universal
algebraic relaxation of fronts propagating into unstable state and
implications for moving boundary approximations}, Phys.\ Rev.\ Lett.\ {\bf 80}, 1650 (1998).

\bibitem{evs2}  
U. Ebert and W. van Saarloos, {\em Front propagation into
unstable states: universal algebraic convergence towards
uniformly pulled fronts}, Physica D {\bf 146}, 1 (2000).


\bibitem{wimreview} W. van Saarloos, {\em Front propagation into
unstable states},  Phys. Rep. {\bf 386}, 29  (2003).


\bibitem{stokes} A. N.\ Stokes, {\em On Two Types of Moving Fronts 
  in Quasilinear Diffusion}, Math.\ Biosciences {\bf 31}, 307
  (1976).   

\bibitem{paquette} G. C.\ Paquette, L.-Y.\ Chen, N.\ Goldenfeld, and Y.\ 
  Oono, {\em Structural Stability and Renormalization Group for
    Propagating Fronts},  Phys.\ Rev.\ Lett.\ {\bf 72}, 76 (1994).  

\bibitem{evs3} U. Ebert and W. van Saarloos, {\em Breakdown of the
standard Perturbation Theory and Moving Boundary
Approximation for "Pulled" Fronts},  Phys. Rep. {\bf 337}, 139 (2000).

\bibitem{fisher} R. A.  Fisher, {\em The wave of advance of
    advantageous genes},  Ann. Eugenics {\bf 7}, 355
(1937). 

\bibitem{kpp} A.~Kolmogoroff, I.~Petrovsky, and N.~Piscounoff, {\em
    Study of the Diffusion Equation with Growth of the Quantity of
    Matter and its Application to a Biology Problem},
  Bulletin de l'universit\'e d'\'etat \`a Moscou, Ser.~int., Section
  A, Vol.~1 (1937).

\bibitem{bramson} M.\ Bramson, {\em Convergence of Solutions of the
    Kolmogorov equation to traveling waves}, Mem.\ Am.\ Math.\ Soc.\
  {\bf 44}, No.\   285 (1983). 

\bibitem{evsp} U. Ebert, W. van Saarloos and L. A. Peletier, {\em
Universal algebraic convergence in time of pulled fronts:
the common mechanism for difference-differential and partial
differential equations},  Euro. Jnl. Appl. Math. {\bf 13},
53 (2002).

\bibitem{bd} E.\ Brunet and B.\ Derrida, {\em Shift of the
velocity of a front due to a cutoff}, Phys.\ Rev.\ E {\bf 56},
  2597 (1997). 

\bibitem{storm}
C. Storm, W. Spruijt, U. Ebert and W. van Saarloos, {\em Universal
  Algebraic Relaxation of Velocity and Phase in Pulled Fronts
  generating Periodic or Chaotic States}, Phys. Rev. E {\bf 61}, R6063
(2000).

\bibitem{dee} G. Dee and J. S. Langer, {\em Propagating Pattern
Selection},  Phys. Rev. Lett. {\bf 50}, 383   (1983).   

\bibitem{vs2} W. van Saarloos, {\em Front Propagation into Unstable
States II: Linear versus Nonlinear Marginal Stability and Rate of
Convergence},  Phys. Rev. A {\bf 39}, 6367 (1989).

\bibitem{collet1} P. Collet and J. P. Eckmann, {\em The Stability of
Modulated  Fronts},  Helv. Phys. Acta {\bf 60}, 969 (1987).

\bibitem{collet2} P. Collet and J. P. Eckmann, {\em Instabilities and
    Fronts in Extended Systems} (Princeton University Press,
Princeton, 1990). 

\bibitem{eckmann} J.-P. Eckmann and C. E. Wayne, {\em Propagating
Fronts and the Center Manifold Theorem},
  Commun. Math. Phys. {\bf 161}, 323 (1994).

\bibitem{collet3} P. Collet and J. P. Eckmann, {\em A rigorous upper
    bound on the propagation speed for the Swift-Hohenberg and related 
    equations}, J. Stat. Phys. {\bf 108}, 1107 (2002). 

\bibitem{deevs} G.\ Dee and W.\ van Saarloos, {\em Bistable Systems
with Propagating Fronts Leading to Pattern Formation},  Phys.\ Rev.\ Lett.\ {\bf 60},
  2641 (1988).  

\bibitem{bert} L. A. Peletier and W. C. Troy,  {\em Spatial Patterns:
    Higher order Models in Physics and Mechanics} (Birk\"auser,
  Boston, 2001).

    
\bibitem{ramses} R. van Zon, H. van Beijeren and Ch. Dellaga, {\em Largest Lyapunov 
exponent for many-particle systems at low densities}, Phys. Rev. Lett. {\bf 80}, 2035 (1998).

\bibitem{nb} K. Nozaki and N. Bekki, {\em Pattern Selection and
Spatio-temporal Transition to Chaos in the Ginzburg-Landau
equation}, Phys. Rev. Lett. {\bf 51}, 2171
  (1983).



\bibitem{vsh} W. van Saarloos and P. C. Hohenberg, {\em Fronts,
Pulses, Sources and Sinks in Generalized Complex Ginzburg-Landau
Equations}, Physica D {\bf 56}, 303  (1992) [Errata: Physica D 
{\bf 61}, 209 (1993)].


\bibitem{abramowitz} M. Abramowitz and I. S. Stegun (Eds.), {\em
    Handbook of Mathematical Functions} (Dover, New York, 1972).



\end{thebibliography}
\end{document}